\documentstyle[floats,aps]{revtex}
\begin{document}
\draft
\title{\large\bf
    Integrable open supersymmetric $U$ model with boundary impurity}


\author{A. Foerster$^1$\thanks{angela@if.ufrgs.br},
K.E. Hibberd$^2$\thanks{keh@cbpf.br},
J.R. Links$^3$\thanks{jrl@maths.uq.edu.au} and
I. Roditi$^4$.\thanks{roditi@cbpf.br}}
\date{\today}
\maketitle
\begin{center}
${}^1$Instituto de Fisica da UFRGS,\\
Av. Bento Goncalves 9500, Porto Alegre - RS, Brazil.\\
\vspace{.5cm}
${}^{2,4}$Centro Brasileiro de Pesquisas Fisicas, \\
Rua Dr. Xavier Sigaud 150, 22290-180, Rio de Janeiro - RJ, Brazil.\\
\vspace{.5cm}
${}^3$Department of Mathematics, \\
The University of Queensland, 4072, Australia. \\

\end{center}

\maketitle

\vspace{10pt}

\begin{abstract}
 An integrable version of the supersymmetric $U$ model with open
 boundary conditions and an impurity situated at one end of the
 chain is introduced. The model is solved through the nested algebraic Bethe ansatz
 method so that the Bethe ansatz equations are obtained.
\end{abstract}




\def\a{\alpha}
\def\b{\beta}
\def\d{\delta}
\def\e{\epsilon}
\def\g{\gamma}
\def\k{\kappa}
\def\l{\lambda}
\def\o{\omega}
\def\t{\theta}
\def\s{\sigma}
\def\D{\Delta}
\def\L{\Lambda}


\def\beq{\begin{equation}}
\def\eeq{\end{equation}}
\def\bea{\begin{eqnarray}}
\def\eea{\end{eqnarray}}
\def\ba{\begin{array}}
\def\ea{\end{array}}
\def\no{\nonumber}
\def\le{\langle}
\def\re{\rangle}
\def\lt{\left}
\def\rt{\right}
\def\dwn{\downarrow}
\def\up{\uparrow}
\def\dag{\dagger}
\def\a{\alpha}
\def\b{\beta}
\newcommand{\sect}[1]{\setcounter{equation}{0}\section{#1}}
\renewcommand{\theequation}{\thesection.\arabic{equation}}
\newcommand{\reff}[1]{eq.~(\ref{#1})}

\vskip.3in

\sect{Introduction\label{int}}

The study of one-dimensional models of correlated electrons in the
context of Luttinger liquids attracts a wealth of activity presently.
Particularly, such models in the presence of defects or impurities are
of particular interest in the effort to unify the results of both
theoretical and experimental investigations. As a result, substantial
research has been devoted to this field from a variety of approaches.

Lee and Toner first considered the problem of an impurity coupled to a 
Luttinger liquid  \cite{LT}
and the Kondo (magnetic impurity) problem has been studied in detail 
using conformal field theory \cite{affl90,AL91,FN94}, perturbation theory 
\cite{K64}, renormalization group \cite{KF92} and Bethe ansatz methods 
\cite{hou,SZ97,A80,TW83}.  Still the problem of strongly correlated electron models with 
impurities remains not well understood and    
recent discrepancies between observed experimental results 
and theoretical predictions \cite{sse} have justified the search for a 
deeper understanding.

Some of the well studied one-dimensional integrable spin models with impurities include the (supersymmetric) t-J model \cite{BEF97,bares,PW97,LF98}
and Heisenberg chain \cite{hou,sse,hpe,schlo,AJ84,LS}. However one discovers that when impurities are introduced into
a periodic lattice the Hamiltonian, though integrable, may contain 
terms that are physically unrealistic  
as the system suffers from an absence of backscattering 
\cite{SZ97,sse,BEF97,bares,LF98,hpe}.

There exists another class of one-dimensional models with impurities.  
These are systems constructed with open boundary conditions where the 
impurity appears on the end of the chain.  
It is desirable to introduce impurities into the system in such a way
that integrability is maintained.  The quantum inverse scattering method (QISM)
allows for such a construction and following Sklyanin's approach
\cite{Skl88} one may obtain an integrable impurity model with open boundary 
conditions.  This has been demonstrated for a number of models with Kondo impurities \cite{wang,stuff,uq,wdhp,bedfrahm,fan,jack}.

This article presents a new boundary impurity for the supersymmetric $U$ model 
\cite{bglz} maintaining integrability. 
Since its discovery, the supersymmetric $U$ model has received 
considerable attention due partly to the fact that the model contains 
a free integrability preserving coupling parameter.
The Bethe ansatz equations and thermodynamic aspects of the model 
on the closed chain were discussed in \cite{them} and an 
anisotropic generalization was presented in \cite{us}.

The boundary impurity we propose for the supersymmetric $U$ model, 
is constructed from an $R$-matrix solution 
of the Yang-Baxter relation.  The solution is obtained from the tensor 
product of one parameter family of typical four-dimensional irreducible 
representations of $gl(2|1)$ characterized by distinct parameters.
This produces boundary terms with a different coupling of the Hubbard interaction, 
compared with that of the bulk Hamiltonian.  In addition,
the boundary term also contains a spin interaction which is 
absent in the bulk interactions.   

The paper is organised as follows.  The construction of the Hamiltonian is 
presented in the next section.  In section III the nested algebraic Bethe ansatz 
solution is presented and concluding remarks follow in section IV.

\sect{The supersymmetric $U$ model with boundary impurity
      \label{for}}

In this section, we introduce the basic ingredients to construct
an integrable impurity model with open boundary conditions through
the QISM.  We begin by writing the $ R $-matrix \cite{g1}
\beq
R^{\a\b}(u) = \frac{u-\alpha-\beta}{-u-\alpha -\beta}
P_1 -  P_2 +
\frac{-u-\alpha-\beta-2}{u - \alpha -\beta -2}
 P_3,
\label{rr}
\eeq
where the $ \alpha, \beta $ arise from the
one-parameter family of four-dimensional representations of
$gl(2|1)$ and the projectors $  P_1 $,
$  P_2 $ and
$P_3 $ are given
explicitly in the appendix. In particular, for $\beta = \alpha$
we recover the $R$-matrix used to derive the usual supersymmetric
$U$ model\cite{bglz}.

This $  R $-matrix acts in the tensor product of
two four-dimensional modules $V_\alpha \otimes V_\beta$
and is a solution to the
${\bf Z}_2$-graded quantum
Yang-Baxter equation
\beq
R^{\a\a}_{12}(u_1-u_2)
R^{\a\b}_{13}(u_1-u_3)
R^{\a\b}_{23}(u_2-u_3) =
R^{\a\b}_{23}(u_2-u_3)
R^{\a\b}_{13}(u_1-u_3)
R^{\a\a}_{12}(u_1-u_2).
\label{qybe}
\eeq
The usual notation is adopted, that is,
$R_{jk}(u)$ acts on $j$-th and $k$-th superspaces and as an
identity on the remaining superspace. Throughout the article
we shall use the graded-tensor product law defined by
\beq
(a\otimes b) (c\otimes d)=(-1)^{[b][c]}\;(ac \otimes bd).
\label{rule}
\eeq
This $R$-matrix also satisfies unitarity
$$R_{12}(u) R_{21} (-u) = \rho(u),$$
and crossing unitarity
$$R^{st_2}_{12}(-u+1) R_{12}^{st_1}(u)={\tilde\rho}(u),$$
where $\rho(u),~~{\tilde\rho}(u)$ are some scalar functions.

In order to deal with open boundary conditions we follow
the necessary supersymmetric 
generalization of Sklyanin's approach \cite{Skl88}
and define the boundary transfer matrix $\tau(u)$ as
\beq
\tau(u)= \mbox{str}_0 \lt( K_+(u)T(u)K_-(u) T^{-1}(-u) \rt).
\label{tau}
\eeq
Here $T(u)$ is the monodromy matrix for a $L$-site chain
\beq
T(u)= R^{\a\a}_{01}(u)
R^{\a\a}_{02}(u) \cdots
 R^{\a\a}_{0(L-1)}(u)
 R^{\a\b}_{0L}(u),
\label{mon}
\eeq
$K_-(u),\,K_+(u)$ both obey the reflection equations
\bea
&& R_{12}(u_1-u_2){K}_-(u_1) R_{21}(u_1+u_2){K}_-(u_2) =
{K}_-(u_2) R_{12}(u_1+u_2)
{K}_-(u_1) R_{21}(u_1-u_2),\no\\
&&R_{21}^{st_1 \; ist_2}(-u_1+u_2) {K^{st_1}_+}(u_1)
R_{12}(-u_1-u_2+\eta)
{K^{ist_2}_+}(u_2)\no\\
&&~~~~~~~~~~~~~~~~~~~= {K^{ist_2}_+}(u_2) R_{21}(-u_1-u_2+\eta)
{K^{st_1}_+}(u_1) R_{12}^{st_1\; ist_2}(-u_1+u_2).
\label{REs}
\eea
Above $\eta$ is the so-called crossing parameter and
$st_i$ stands for the supertransposition taken in the $i$-th space,
whereas $ist_i$ is the inverse operation of $st_i$.  Here we will choose $K_-(u),\,K_+(u)$ to be
both equal to the identity matrix.
Note that in the expression for the monodromy matrix
$T(u)$ (\ref{mon}) the presence of an $R$-matrix intertwining
between two different four-dimensional representations
of $gl(2|1)$ characterized by $\alpha$ and $\beta$.
The corresponding transfer matrix will be associated
to a supersymmetric $U$ model with an impurity
in the boundary.

Using the Yang-Baxter algebra (\ref{qybe}), together with the
reflection equations (\ref{REs}), it can be shown that the model
is integrable, that is
\beq
[\tau(u_1),\tau(u_2)] = 0.
\label{int}
\eeq
From this transfer matrix one may write the Hamiltonian for an open
supersymmetric $U$ model with a boundary impurity through the
second derivative of $\tau(u)$ with respect to the spectral parameter u, at
$u=0$. This is the procedure adopted when
the supertrace of $K_+(u)$ is null \cite{hh}.  The Hamiltonian is given by
\bea
H&=&-\frac {q^{\a+1}-q^{-\a-1} }{ \ln q} H^R,\nonumber\\
H^R &=& \frac { \tau''(0)} {4(A_1+2B_1)}\nonumber \\
   & = &\sum_{i=1}^{L-1} H_{i,i+1}+\frac 12 {\stackrel 1K}{}'_-(0)\nonumber \\
&&+\frac 1{2(A_1+2B_1)}
\left[ \mbox{str}_0\left(\stackrel 0K_+(0) G_{L0}\right)\right.
\left. +2~\mbox{str}_0\left({\stackrel 0K}{}'_+(0)H^R_{L0}\right)+\mbox{str}_0\left(\stackrel 0K_+(0)(H^R_{L0})^2\right)\right],\label{impham}
\eea
where
\begin{eqnarray*}
A_1 &=& \mbox{str}_0 K'_+(0), ~~~B_1=\mbox{str}_0\left(\stackrel 0K_+(0) H^R_{L0}\right),~~~
H^R_{i,i+1}= P_{i,i+1} R'_{i,i+1}(0),~~~G_{i,i+1}=P_{i,i+1} R''_{i,i+1}(0).\nonumber \\
\end{eqnarray*}
In the above, $P_{i,i+1}$ denotes the graded permutation operator acting on quantum spaces $i$ and $(i+1)$.
With our choice for the boundary K matrices equal to the identity, the Hamiltonian (\ref{impham}) simplifies to
\bea
H^R&=& \frac { \tau''(0)} {8~ \mbox{str}_0 (H^R_{L0})  }= \sum_{i=1}^{L-1} H_{i,i+1} +\frac 1{4~\mbox{str}_0 (H^R_{L0})}
\left[ \mbox{str}_0 (G_{L0}) +\mbox{str}_0\left((H^R_{L0})^2\right)\right]. \label{ourham}
\eea
In view of the grading, the basis vectors of the module $V$ can
be identified with the eletronic states as follows
\bea
|1\rangle \equiv |\uparrow \downarrow \rangle = c^\dag_\downarrow c^\dag_\uparrow|0\rangle \ ,
\ |2\rangle \equiv |\uparrow \rangle =c^\dag_\uparrow|0\rangle \ , \
|3\rangle \equiv |\downarrow \rangle = c^\dag_\downarrow |0\rangle \ ,
\ |4\rangle \equiv | 0\rangle.  \label{states}
\eea
We adopt the notation that $c_{i,\uparrow}^{\dagger}$ ($c_{i,\downarrow}^{\dagger}$) are spin up annihilation
(spin down creation) operators. 
Also let $n_{i,\s}=c_{i,\s}^\dagger c_{i,\s}$ denote the number operator for
electrons with spin $\s$ on site $i$ and let the total number of electrons be $
n_i=n_{i,\uparrow}+
n_{i,\downarrow}$. The spin operators $S\,,~S^\dagger\,,~
S^z$,
\begin{eqnarray*}
S_i=c_{i,\uparrow}^\dagger c_{i,\downarrow}\,,~~~
  S_i^\dagger=c_{i,\downarrow}^\dagger c_{i,\uparrow}\,,~~~
  S_i^z=\frac{1}{2}(n_{i,\downarrow}-n_{i,\uparrow})\,,
\end{eqnarray*}
form an $sl(2)$ algebra.
These operators commute with the Hamiltonian of the supersymmetric $U$ model 
which is given by
\begin{eqnarray*}
H_{i,i+1}&=& -\sum_{\sigma=\uparrow,\downarrow}(c_{i,\sigma}^\dagger c_{i+1,\sigma}
 +\mbox{h.c.})\left(\frac {1+\alpha}{\alpha}\right)^{1/2(n_{i,-\sigma}+n_{i+1,-\sigma})}
+\frac 1{\alpha}\left [ n_{i,\uparrow}n_{i,\downarrow} +n_{i+1,\uparrow} n_{i+1,\downarrow}\right ]\nonumber\\
& &+\frac{1}{\alpha} (c_{i,\sigma}^\dagger c_{i,-\sigma} ^\dagger c_{i+1,-\sigma} c_{i+1,\sigma}+\mbox{ h.c.})
 +(n_i+n_{i+1}).
\end{eqnarray*}

The impurity Hamiltonian constructed from (\ref{ourham}) is found to be 
\bea
H=\sum_{i=1}^{L-1}H_{i,i+1}+ H^\beta_{L,0}, \label{ham1}
\eea
where the boundary term $H^\beta_{L,0}$ is given 
\begin{eqnarray}
H^\beta_{L,0} &=& l\{t+\sum_{\sigma} ( c_{L-1,\sigma}^\dag c_{L,\sigma}+h.c.) 
[ h+ d(\a,\b) n_{L-1,-\sigma} (n_{L,-\sigma} -1)+d(\b,\a) (n_{L-1,-\sigma} -1) n_{L,-\sigma}\nonumber \\
&&+ m n_{L-1,-\sigma} n_{L,-\sigma} ]-\frac e2(c^\dag_{L-1,\s} c_{L-1,-\s}^\dag c_{L,-\s} c_{L,\s}  +h.c.)
-f(S^\dag_{L-1} S_L+h.c.)+f \sum_{\sigma}  n_{L-1,\sigma}n_{L,-\sigma} \nonumber \\
&&- (\a+\b)^2 [ (1+\b) n_{L-1} +(1+\a) n_L ]
-4 [ \b(1+\b) n_{L-1,\up}n_{L-1,\dwn} 
+ \a(1+\a) n_{L,\up}n_{L,\dwn}]\},\label{boundary}
\end{eqnarray}
with
\begin{eqnarray*}
m &=& \sqrt{\a\b}(2+\a+\b)^2 -(\a+\b)^2\sqrt{(1+\a)(1+\b)} ,~~~
d(\a,\b) =-\sqrt{1+\b} [4\sqrt{\a}+(\a+\b)^2(\sqrt{\a} - \sqrt{1+\a} )],
\\
f &=&(\a-\b)^2,~~~
e=4\sqrt{\a\b(1+\a)(1+\b)}, ~~~
t=4 (\a+\b)^2(2+\a+\b),\\
h&=&(\a+\b)^2\sqrt{(1+\a)(1+\b)} ,~~~
l= \frac{-4(\b+1)}{(\a+\b)^2(2+\a+\b)^2}.
\end{eqnarray*}

At this point a few interesting characteristics of the model should be pointed
out.  In contrast to the usual
supersymmetric $U$ model, one may observe that the above Hamiltonian (\ref{ham1}) contains an extra 
spin interaction term from the boundary contribution.  Also 
the free parameter $\alpha$ appears on the boundary as well as in the bulk terms of the Hamiltonian, in contrast 
the Kondo impurity model presented in \cite{jack}.
Furthermore, in the limit $\beta \rightarrow \alpha $ we recover
the usual supersymmetric $U$ model \cite{bglz}.

\sect{The Bethe ansatz solutions \label{bethe}}

In order to find the Bethe ansatz equations of the Hamiltonian (\ref{ham1}),
we use Sklyanin's generalized Bethe ansatz method to treat open boundary
conditions together with Babujian and Tsvelick's \cite{BT}
approach for higher spin chains. Therefore, we
introduce the following doubled auxiliary monodromy
matrix 
\beq
\hat{\cal{U}}(u)= \lt( \hat{T}(u)\hat{K}_-(u) {\hat{T}}^{-1}(-u) \rt),
\label{dobmonaux}
\eeq
where we will chose $\hat{K}_-(u)$ to be the identity matrix and
$\hat{T}(u)$ is the monodromy matrix for a $L$-site chain
\beq
\hat{T}(u)= R^{\a}_{01}(u)
 R^{\a}_{02}(u) \cdots
 R^{\a}_{0(L-1)}(u)
 R^{\b}_{0L}(u).
\label{monaux}
\eeq
Above the matrix $ R^{\a}(u)$ acts on $W
\otimes V_{\alpha } $, where $W$ is a three-dimensional module and it
is given by \cite{us}
\begin{equation}
\label{rstar}
R^{\a}(u)=
\pmatrix{1&0&0&0&|&  0&0&0&0&|&  0&0&0&0\cr
 0&1&0&0&|&0&0&0&0&|&  0&0&0&0\cr
 0&0&b^*&0&|&  0&c^*&0&0&|& d^*&0&0&0\cr
 0&0&0&b^*&|&  0&0&0&0&|& 0&e^*&0&0\cr
 -&-&-&-& &  -&-&-&-& &  -&-&-&-\cr
 0&0&0&0&|&1&0&0&0&|&  0&0&0&0\cr
 0&0&c^*&0&|&  0&b^*&0&0&|&-d^*&0&0&0\cr
 0&0&0&0&|&  0&0&1&0&|&  0&0&0&0\cr
 0&0&0&0&|&  0&0&0&b^*&|&  0&0&e^*&0\cr
 -&-&-&-& &  -&-&-&-& &  -&-&-&-\cr
 0&0&-d^*&0&|&  0&d^*&0&0&|&w^*&0&0&0\cr
 0&0&0&-e^*&|&  0&0&0&0&|&  0&f^*&0&0\cr
 0&0&0&0&|&  0&0&0&-e^*&|&  0&0&f^*&0\cr
 0&0&0&0&|&  0&0&0&0&|&  0&0&0&g^*\cr}  ,
\end{equation}
with
\begin{eqnarray}
b^* &=& 1 + 1/u ,\quad
c^* = 1/u ,\quad
d^* =  \sqrt{\alpha}/u  ,\quad
e^* =  \sqrt{\alpha + 1}/u  ,\quad \nonumber \\
f^* &=& 1 +  (\alpha + 1)/u ,\quad
w^* = 1 +  \alpha /u , \quad
g^* = 1 +  ( \alpha + 2 )/u.
\label{elem}
\end{eqnarray}

The doubled auxiliary monodromy matrix (\ref{dobmonaux})
acts on $W \otimes {V_\alpha}^{L-1} \otimes V_\beta $ and can be represented by the matrix
\begin{eqnarray}
\hat{\cal{U}}(u)=
\left( \begin{array}{ccc}
{\cal A}_{11}(u) & {\cal A}_{12}(u) & {\cal B}_1(u)\\
{\cal A}_{21}(u) & {\cal A}_{22}(u) & {\cal B}_2(u)\\
{\cal C}_1(u)    & {\cal C}_2 (u)   & {\cal D}(u)
\end{array} \right).\label{u1}
\end{eqnarray}
It is possible to show that it fulfills a modified
Yang-Baxter equation of the form
\begin{equation}
\footnotesize{
R_{1 2}(u_1 - u_2) \,
\hat{\cal U}(u_1) \,
R_{2 1}(u_1 + u_2) \,
\hat{\cal U}(u_2) \, = \,
\hat{\cal U}(u_2) \,
R_{1 2}(u_1 + u_2) \,
\hat{\cal U}(u_1) \,
R_{2 1}(u_1 - u_2) },
\label{genyba}
\end{equation}
where $R(u)$ is the usual $t-J$ $R$-matrix acting on
$W \otimes W$\cite{EK92,FK92}
\beq
R(u)=
\pmatrix{
u-1&0&0 & |&     0&0&0 &   |&     0&0&0\cr
0&u&0   & |&  -1&0&0 &      |&  0&0&0 \cr
0&0&u&  |&     0&0&0&      |& 1&0&0\cr
-&-&-& &  -&-&-& &  -&-&-\cr
0&-1&0 &  |&  u&0&0&   |&    0&0&0\cr
0&0&0&   |&   0&u-1&0& |&   0&0&0\cr
0&0&0&     |& 0&0&u&    |&  0&1&0\cr
-&-&-& &  -&-&-& &  -&-&-\cr
0&0&-1&  |&  0&0&0&  |&    u&0&0\cr
0&0&0&  |&   0&0&-1&  |&    0&u&0\cr
0&0&0&    |& 0&0&0&   |&   0&0&u+1\cr}.\label{r1}
\eeq
From the doubled auxiliary monodromy matrix (\ref{dobmonaux})
we define an auxiliary transfer matrix
\beq
\hat{\tau}(u)= \mbox{str}_0 \lt( \hat{K}_+(u) \hat{\cal{U}}(u) \rt),
\label{tauaux}
\eeq
where $\hat{K}_+(u) $ is chosen to be the
diagonal matrix with elements $\{ -1,-1,1 \}$.

It can be shown that this auxiliary transfer matrix $ \hat{\tau}(u)$ (\ref{tauaux})
commutes with the transfer matrix $\tau(u)$ (\ref{tau}), which
means that they have a common set of eigenvectors.  These vectors
will be determined by applying the algebraic
Bethe ansatz to $\hat{\tau}(u)$.
For this purpose we begin by writing the Bethe vector
$|\Omega \rangle $ (according to the first-level Bethe ansatz) as
\beq
|\Omega \rangle = {\cal C}_{i_1}(u_1) \cdots {\cal
C}_{i_N}(u_N)|\Psi\rangle F^{i_1\cdots i_N}.
\label{bv}
\eeq
$|\Psi\rangle $ is the pseudovacuum and the coefficients $F^{i_1\cdots i_N}$ will be determined
later by the second level Bethe ansatz. The action of the elements of $\hat{\cal U}(u)$
(\ref{u1}) on $|\Psi\rangle $ is given by
\bea
{\cal D}(u)|\Psi\rangle&=&
\biggl(\frac {u+\alpha +2}{-u + \alpha +2}\biggr)^{L-1}
\frac {(u+\beta +2)}{(-u + \beta +2)}
|\Psi\rangle,\no\\
{\cal B}_d(u)|\Psi\rangle&=&0,~~~~~~
{\cal C}_d(u)|\Psi\rangle\neq 0,\no\\
{\cal A}_{db}(u)|\Psi\rangle&=&
\biggl(\frac {u+1}{-u}\biggr)^L
\biggl(\frac {-u+\alpha +1}{-u + \alpha +2}\biggr)^{L-1}
\frac {(-u+\beta +1)}{(-u + \beta +2)}
|\Psi\rangle,~~~~~(b \neq d)\no\\
{\cal A}_{dd}(u)|\Psi\rangle&=&
\biggl(\frac {u+1}{-u}\biggr)^L
\biggl(\frac {-u+\alpha +1}{-u + \alpha +2}\biggr)^{L-1}
\frac {(-u+\beta +1)}{(-u + \beta +2)}
\lt(\frac{2u}{2u+1}\rt)|\Psi\rangle \nonumber \\
&&+\frac{1}{2u+1}
\biggl(\frac {u+\alpha +2}{-u + \alpha +2}\biggr)^{L-1}
\frac {(u+\beta +2)}{(-u + \beta +2)}
|\Psi\rangle.
\eea
Using the Yang-Baxter relation (\ref{genyba}) together with the
expressions for the matrices $\phantom{0}_{W W}R(u)$ (\ref{r1}) and
$\hat{\cal U}(u)$ (\ref{u1}), we can obtain the commutation relations between
${\cal D}(u),\, {\cal A}_{db}(u) $ and ${\cal C}_d(u)$.
These are complicated relations but they will simplify considerably if instead of
using ${\cal A}_{db}(u)$ we work with $\tilde{{\cal A}}_{db}(u)$ according to the transformation
\beq
{\tilde{\cal A}}_{bd}(u) = {\cal A} _{bd}(u) - \frac {1}{2u+1}
\delta _{bd}{\cal D}(u).  \nonumber
\eeq
This transformation will allow us to more easily recognize the ``wanted" and ``unwanted" terms.
More explicitly we have
\bea
{\tilde{\cal A}}_{bd}(u_1){\cal C}_c(u_2)&=&\frac {(u_1-u_2-1)(u_1+u_2)}
{(u_1-u_2)(u_1+u_2+1)}~r^{eb}_{gh}(u_1+u_2+1)~r^{ih}_{cd}(u_1-u_2)~
{\cal C}_e(u_2){\tilde{\cal A}}_{gi}(u_1)\no\\
& &-\frac {4u_1u_2}{(u_1+u_2+1)(2u_1+1)(2u_2+1)}~r^{gb}_{cd}(2u_1+1)~
{\cal C}_g(u_1) {\cal D}(u_2)  \no\\
& & +\frac {2u_1}{(u_1-u_2)(2u_1+1)}~
r^{gb}_{id}(2u_1+1) ~{\cal C}_g (u_1) {\tilde{\cal A}}_{ic}(u_2),\label
{cr1}\\
{\cal D}(u_1){\cal C}_b(u_2)
&=&\frac {(u_1-u_2-1)(u_1+u_2)}
{(u_1-u_2)(u_1+u_2+1)}
{\cal C}_b(u_2){\cal D}(u_1)+\frac {2u_2}{(u_1-u_2)(2u_2+1)}
{\cal C}_b(u_1){\cal D}(u_2)\no\\
& &  -\frac {1}{u_1+u_2+1}{\cal C}_d(u_1){\check {\cal A}}_{db}(u_2).
\eea
Here the matrix $r(u)$, which satisfies the Yang-Baxter
equation, takes the form,
\beq
r^{bb}_{bb}(u)=1,~~~~~r^{bd}_{bd}(u)=-\frac
{1}{u-1},~~~~~r^{bd}_{db}(u)=\frac {u}{u-1},~~~(b \neq d,~~b,d = 1,2).
\eeq
By applying the transfer matrix $\hat{\tau}(u)$ (\ref{tauaux}) to the
Bethe vector $|\Omega \rangle $ (\ref{bv}) we can solve the eigenvalue problem
$$\hat{\tau}(u) |\Omega \rangle =\hat{\Lambda}(u) |\Omega \rangle.$$ 
We find the eigenvalue to be 
\bea
\hat{\Lambda}(u)&=& \frac {2u-1}{2u+1}
\biggl(\frac {u+\alpha +2}{-u + \alpha +2}\biggr)^{L-1}
\left(\frac {u+\beta +2}{-u + \beta +2}\right)
\prod ^N_{j=1} \frac {(u+u_j)(u-u_j-1)}{(u-u_j)(u+u_j+1)}\no\\
& &-\frac {2u}{2u+1}
\biggl(\frac {u+1}{-u}\biggr)^L
\biggl(\frac {-u+\alpha +1}{-u + \alpha +2}\biggr)^{L-1}
\left(\frac {-u+\beta +1}{-u + \beta +2}\right)
\prod ^N_{j=1} \frac {(u+u_j)(u-u_j-1)}{(u-u_j)(u+u_j+1)}
{\hat{\Lambda}}^{(1)}(u;\{u_i\}),
\eea
provided the parameters $\{ u_j\}$ satisfy a first Bethe ansatz equation given by
\bea
&&\biggl(\frac {-u_j}{u_j+1}\biggr)^{L}
\biggl(\frac {u_j + \alpha + 2}{-u_j +\alpha +1}\biggr)^{L-1}
\left( \frac {u_j + \beta + 2}{-u_j +\beta +1}\right)=\no\\
&&\mbox{\hspace{.5cm}}\prod _{m=1}^M \frac
{(u_j-v_m+\frac {3}{2})(u_j+v_m+\frac {1}{2})}
{(u_j-v_m+\frac {1}{2})(u_j+v_m-\frac {1}{2})} 
  + \frac {2u_j+1}{2u_j-1}
\prod _{k=1}^N \frac {(u_j-u_k)(u_j+u_k+1)}
{(u_j-u_k-1)(u_j+u_k)}
\prod ^{M}_{m=1} \frac {(u_j-v_m-\frac {1}{2})(u_j+v_m-\frac {3}{2})}
{(u_j-v_m+\frac {1}{2})(u_j+v_m-\frac {1}{2})}
. \label {bethe11}
\eea
Above, ${\hat{\Lambda}}^{(1)}(u;\{u_i\})$ is the eigenvalue of the transfer
matrix ${\hat{\tau}}^{(1)}(u)$ for the reduced problem which arises out of
the $r(u)$ matrices from the first term in the right hand side of (\ref {cr1}).
The reduced transfer matrix
$\hat{ \tau} ^{(1)}(u)$ may be recognized as that of the $N$-site inhomogeneous
XXX  spin-$\frac {1}{2}$ open chain
and may be diagonalized following Ref.\cite {Skl88} with the reduced 
boundary K matrices are equal to the identity.  
We find the eigenvalue of the transfer matrix  $ \hat{\tau} ^{(1)}(u)$ to be given by
\bea
\hat{\Lambda}^{(1)}(u;\{ u_j \}) &=&
 \frac {2u-1}{2u} \prod _{m=1}^M \frac
{(u-v_m+\frac {3}{2})(u+v_m+\frac {1}{2})}
{(u-v_m+\frac {1}{2})(u+v_m-\frac {1}{2})}\nonumber \\
&& + \frac {2u+1}{2u}
\prod _{j=1}^N \frac {(u-u_j)(u+u_j+1)}
{(u-u_j-1)(u+u_j)}
\prod ^{M}_{m=1} \frac {(u-v_m-\frac {1}{2})(u+v_m-\frac {3}{2})}
{(u-v_m+\frac {1}{2})(u+v_m-\frac {1}{2})},\no
\eea
provided the parameters $\{ v_m \}$ satisfy a second Bethe ansatz equation given by
\beq
\prod ^N_{j=1} \frac {(v_m-u_j-\frac {3}{2})(v_m+u_j-\frac {1}{2})}
{(v_m-u_j-\frac {1}{2})(v_m+u_j+\frac {1}{2})}
=\prod ^{M}_{\stackrel {k=1}{k \neq m}} \frac {(v_m-v_k-1)
(v_m+v_k-2)} {(v_m-v_k+1)
(v_m+v_k)}. \label {bethe21}
\eeq

\sect{Conclusion}

We have presented the supersymmetric $U$ model with a new type of boundary impurity. 
The model is constructed with open boundary conditions and formulated within a modified QISM 
so that integrability is not lost.
The impurity is introduced through an $R$-matrix solution of the Yang-Baxter equation,
built from the tensor product of two different four-dimensional representations 
of $gl(2|1)$ which are characterized by different parameters.
The resulting Hamiltonian (\ref{ham1}) contains an extra spin interaction term that does not 
appear in the usual supersymmetric $U$ model.  Another feature of this impurity model is that
the free parameter of the bulk appears in the boundary terms, in contrast to
the Kondo impurity model presented in \cite{jack}.  In the limit $\beta \rightarrow \alpha $ 
we recover the supersymmetric $U$ model \cite{bglz} on an open chain.
In this work, we have also derived the Bethe ansatz equations using the 
nested algebraic Bethe ansatz method, which was appropriately modified to deal 
with a graded underlying algebra and open boundary conditions.

\vspace{0.55cm}

ACKNOWLEDGEMENTS

A.F, K.H and I.R would like to thank CNPq (Conselho Nacional de
Desenvolvimento Cient\'{\i}fico e Tecnol\'ogico) for financial support.
J.L. is supported by an Australian Research Council Postdoctoral
Fellowship.

\section*{Appendix:  Projection operators for the $ R^{\a\b}$-matrix}

We present the projectors for the tensor product ${V_\alpha\otimes V_\beta}$.  For details on the construction of $R$-matrices from these projectors, the reader is referred to \cite{g1}.

The tensor product basis is constructed from (\ref{states}) and the tensor product decomposition is ${V_\alpha \otimes V_\beta}=V(0,0,|\alpha+\beta)\oplus V(-1,-1,|\alpha+\beta+2)\oplus V(0,-1,|\alpha+\beta+1)$.  It is not necessary to calculate the three
 projectors as one may use the relation $Ident= \sum_i P_i$.
The simplest projectors to calculate are $ P_1 $ and
$P_3 $ which are given as follows
\begin{equation}
\label{p1}
 P_1 =
\pmatrix{
 1&0&0&0&  0&0&0&0 & 0&0&0&0&  0&0&0&0\cr
 0&\b d_2&0&0&  -\sqrt{\a\b} d_2&0&0&0 & 0&0&0&0&  0&0&0&0\cr
 0&0&\b d_2&0&  0&0&0&0   &-\sqrt{\a\b} d_2 &0&0&0&  0&0&0&0\cr
 0&0&0&\b g_0 d_1 &  0&0&-f_2 d_1&0 & 0&f_2 d_1&0&0&  f_0 d_1&0&0&0\cr
 0&\sqrt{\a\b} d_2&0&0&  \a d_2 &0&0&0 & 0&0&0&0&  0&0&0&0\cr
 0&0&0&0&  0&0&0&0 & 0&0&0&0&  0&0&0&0\cr
 0&0&0&f_2 d_1&  0&0&\b\a d_1&0 & 0&-\b\a d_1&0&0&  f_1 d_1&0&0&0\cr
 0&0&0&0&  0&0&0&0 & 0&0&0&0&  0&0&0&0\cr
 0&0&\sqrt{\a\b} d_2&0&  0&0&0&0 &  \a d_2&0&0&0&  0&0&0&0\cr
 0&0&0&-f_2 d_1&  0&0&-\b\a d_1&0 & 0&\b\a d_1&0&0&  -f_1 d_1&0&0&0\cr
 0&0&0&0&  0&0&0&0 & 0&0&0&0&  0&0&0&0\cr
 0&0&0&0&  0&0&0&0 & 0&0&0&0&  0&0&0&0\cr
 0&0&0&f_0 d_1&  0&0&-f_1 d_1&0 & 0&f_1 d_1&0&0&  \a c_0 d_1&0&0&0\cr
 0&0&0&0&  0&0&0&0 & 0&0&0&0&  0&0&0&0\cr
 0&0&0&0&  0&0&0&0 & 0&0&0&0&  0&0&0&0\cr
 0&0&0&0&  0&0&0&0 & 0&0&0&0&  0&0&0&0\cr}  ,\nonumber
\end{equation}
with 
\begin{eqnarray}
d_1&=& 1/[\a(\a+1)+2\a\b+\b(\b+1)]  ,~~~ d_2=1/(\a+\b) ,\nonumber \\
f_0&=&\sqrt{\a\b(\a+1)(\b+1) }  ,~~~ f_1=\a\sqrt{\b(\a+1)}    ,~~~f_2=\b\sqrt{\a(\b+1)}.\nonumber 
\end{eqnarray}

\begin{equation}
\label{p3}
 P_3=
\pmatrix{
 0&0&0&0&           0&0&0&0 &     0&0&0&0& 0&0&0&0\cr
 0&0&0&0&          0&0&0&0 &     0&0&0&0&  0&0&0&0\cr
 0&0&0&0&            0&0&0&0 &     0&0&0&0&  0&0&0&0\cr
 0&0&0&\a c_0g_1&  0&0&c_1 g_1&0 & 0&-c_1 g_1&0&0&  f_0 g_1&0&0&0\cr
 0&0&0&0&            0&0&0&0 &                 0&0&0&0&  0&0&0&0\cr
 0&0&0&0&            0&0&0&0 &                 0&0&0&0&  0&0&0&0\cr
 0&0&0&-c_1 g_1&       0&0&c_3 g_1&0&   0&-c_3g_1&0&0&  -c_2g_1&0&0&0\cr
 0&0&0&0&            0&0&0&c_0g_2 & 0&0&0&0&  0&\sqrt{c_3}g_2&0&0\cr
 0&0&0&0&           0&0&0&0 &                  0&0&0&0&  0&0&0&0\cr 
 0&0&0&c_1g_1&       0&0&-c_3g_1&0 &   0&c_3g_1&0&0&  c_2g_1&0&0&0\cr
 0&0&0&0&           0&0&0&0 &                  0&0&0&0&  0&0&0&0\cr
 0&0&0&0&           0&0&0&0 &                0&0&0&c_0g_2&  0&0&\sqrt{c_3}g_2&0\cr
 0&0&0&f_0g_1&        0&0&c_2g_1&0&            0&-c_2g_1&0&0&  \b g_0g_1&0&0&0\cr
 0&0&0&0&           0&0&0&-\sqrt{c_3}g_2 &         0&0&0&0&  0&g_0g_2&0&0\cr
 0&0&0&0&           0&0&0&0 &               0&0&0&-\sqrt{c_3}g_2&  0&0&g_0g_2&0\cr
 0&0&0&0&           0&0&0&0 & 0&0&0&0&    0&0&0&1\cr}  ,\nonumber
\end{equation}
with
\begin{eqnarray}
c_0 &=& \a+1,~~~
c_1=(\a+1)\sqrt{\a(\b+1)}   ,~~~c_2=(\b+1)\sqrt{\b(\a+1)} ,~~~c_3=(\a+1)(\b+1),   \nonumber \\
g_0&=&\b+1,~~~g_1=1/[\a(\a+1)+2(1+\a)(1+\b)+\b(\b+1)],~~~g_2=1/(\a+\b+2).\nonumber
\end{eqnarray}


\begin{thebibliography}{99}
\bibitem{LT} D. H. Lee and J. Toner, Phys. Rev. Lett.
{\bf 69}, (1992) 3378.

\bibitem{affl90} I. Affleck, Nucl. Phys. {\bf B336}, (1990) 517.

\bibitem{AL91} I. Affleck and A. W. Ludwig, 
Phys. Rev. Lett. {\bf 67}, (1991) 161; Nucl. Phys. {\bf B360},
(1991) 641.

\bibitem{FN94} A. Furusaki and N. Nagaosa, Phys. Rev. Lett.
{\bf 72}, (1994) 892.

\bibitem{K64} J. Kondo, Prog. Theor. Phys. {\bf 32},  (1964) 37.
\bibitem{KF92} C. L. Kane and M. P. A. Fisher, Phys. Rev. Lett. {\bf
68}, (1992) 1220.

\bibitem{hou} B. Hou, K. Shi, R. Yue, S. Zhao, 
``Bethe Ansatz for the spin-1 XXX chain with two impurities," cond-mat/9910383;\\
B. Hou, K. Shi, R. Yue, S. Zhao,
``Integrability of the Heisenberg chains with boundary impurities and their Bethe ansatz," cond-mat/9910384.

\bibitem{SZ97} P. Schlottmann and A.A. Zvyagin, Phys. Rev. {\bf B56}, (1997) 13989.
\bibitem{A80} N. Andrei, Phys. Rev. Lett. {\bf 45}, (1980) 379.
\bibitem{TW83} A.M. Tsvelick, P.B. Wiegmann, Adv. Phys. {\bf32} (1983) 453.
\bibitem{sse} P. Schmitteckert, P. Schwab, U. Eckern, Europhys. Lett. {\bf 30}, (1995) 543.
\bibitem{BEF97} G. Bed\"{u}rftig, F. H. L. Essler, and H. Frahm, Nucl.
Phys. {\bf B498}, (1997) 697;\\
A. Foerster, J. Links and A. Tonel, Nucl. Phys. {\bf B552}, (1999) 707.

\bibitem{bares} P.-A. Bares, ``Exact results for a one-dimensional t-J
model with impurities," cond-mat/9412011.
\bibitem{PW97} Z.-N. Hu, F.-C. Pu and Y. Wang, J. Phys {\bf A31}, (1998) 5241;\\
Z.-N. Hu and F.-C. Pu, Nuclear Phys. {\bf B546}, [FS] (1999) 691.
\bibitem{LF98} J. Links and A. Foerster, J. Phys. {\bf A32} (1999) 147;\\
J. Abad and M. Rios, J. Phys. {\bf A32} (1999) 3535.
\bibitem{hpe} H.-P. Eckle, A. Punnoose, R.A. Romer, Europhys. Lett. {\bf 39}, (1997) 293.
\bibitem{schlo} P. Schlottmann, J. Phys.: Condens. Matter {\bf 3}, (1991) 6617.
\bibitem{AJ84} N. Andrei and H. Johanesson, Phys. Lett. {\bf A100}, (1984) 108.
\bibitem{LS} K. Lee and  P. Schlottmann, Phys. Rev. {\bf B37}, (1988) 379. 
\bibitem{Skl88} E. K. Sklyanin, J. Phys. {\bf A21}, (1988) 2375.
\bibitem{wang} Y. Wang, Phys. Rev. {\bf B56} (1997) 14045.
\bibitem{stuff}H. Fan, M. Wadati and R.-H. Yue, ``Boundary Kondo impurities in the generalized supersymmetric t-J model," cond-mat/9906409. 

\bibitem{uq} H.-Q. Zhou and M. D. Gould,  Phys. Lett. {\bf A251}, (1999) 279;\\
H.-Q. Zhou, X.-Y. Ge, M.D. Gould, J. Phys. {\bf A32}, (1999) 5383;\\
H.-Q. Zhou, X.-Yu Ge, J. Links and M. D. Gould,  Nucl. Phys. {\bf B546}, (1999) 779.

\bibitem{wdhp} Y. Wang, J. Dai, Z. Hu, F-C. Pu, Phys. Rev. Lett. {\bf 79}, (1997) 1901.
\bibitem{bedfrahm} G. Bed\"urftig and H. Frahm, J. Phys. {\bf A32} (1999) 4585.
\bibitem{fan} H. Fan and M. Wadati, "Exact diagonalization of the generalized
supersymmetric t-J model with boundaries", cond-mat/9906083.

\bibitem{jack} H.-Q. Zhou, X.-Y. Ge, M.D. Gould, J. Phys. {\bf A32}, (1999) L137.

\bibitem{bglz} A. J. Bracken, M. D. Gould, J. R. Links and Y.-Z. Zhang, Phys. Rev. Lett. {\bf 74}, (1995) 2768.

\bibitem{them} G. Bedurftig, H. Frahm, J. Phys. {\bf A28}, (1995) 4453;\\
 P.B. Ramos, M.J. Martins, Nucl. Phys. {\bf B474}, (1996) 678; \\
M.P. Pfannm\"uller, H. Frahm, Nucl. Phys. {\bf B479}, (1996) 575.

\bibitem{us} M.D. Gould, K.E. Hibberd, J.R. Links, Y.-Z. Zhang, Phys. Lett. {\bf A212} (1996) 156;\\
  K.E. Hibberd, M.D. Gould, J.R. Links, Phys. Rev. {\bf B 54}, (1996) 8430.


\bibitem{g1} A. Bracken, M. Gould, Y.-Z. Zhang and G. Delius, J. Phys. {\bf A27}, (1994) 6551.


\bibitem{hh}  A. Bracken, X.-Y. Ge, Y.-Z. Zhang, H.-Q. Zhou,  Nucl. Phys. {\bf B516}, (1998) 588.

\bibitem{ZZ97} Y.-Z. Zhang, H.-Q. Zhou,  Phys. Rev. {\bf B58}, (1998) 51.
\bibitem{BT} H.M. Babujian and A.M. Tsvelick, Nucl. Phys. {\bf B265},
(1986) 24. 
\bibitem{EK92} F.H.L. Essler and V. E. Korepin, Phys. Rev. {\bf B46}, (1992) 9147.

\bibitem{FK92} A. Foerster and M. Karowski, Phys. Rev. {\bf B46}, 
(1992) 9234; \\
 A. Foerster and M. Karowski, Nucl. Phys. {\bf B396},  (1993) 611.




\end{thebibliography}
\end{document}